# Performance of the ALICE muon trigger system in pp and Pb–Pb collisions at the LHC

**Gabriele Gaetano Fronzé**[a], **for the ALICE collaboration**

[a] *INFN and University of Turin,*
*Via Pietro Giuria 1, 10125, Torino , Italy*
*E-mail*: gfronze@cern.ch

ABSTRACT: The ALICE muon spectrometer studies the production of quarkonia and open heavy-flavour particles. It is equipped with a Trigger System composed of Resistive Plate Chambers which, by applying a transverse-momentum-based muon selection, minimises the background from light-hadron decays.

The system has been continuously taking data throughout the LHC Run I; it has undergone maintenance and consolidation operations during the LHC shutdown period 1. In the first year of the LHC Run II, the system, fully recommissioned, has participated in data taking in pp and Pb–Pb collisions. The performance of the system throughout the last data-taking period is presented.

## Contents

## 1. The ALICE experiment

ALICE (A Large Ion Collider Experiment) is one of the four large experiments at the Large Hadron Collider. It is specialized in the study of ultra-relativistic heavy ion collisions in order to detect and characterise a new state of matter: the Quark–Gluon Plasma (QGP [1]). The production of heavy quarkonium (and open heavy flavour) states is affected by the QGP formation.

The ALICE muon spectrometer [3] detects the decay products of these states in their $\mu^+\mu^-$ channel. The spectrometer acceptance covers the pseudorapidity interval $-4.0 \leq \eta \leq -2.5$ and allows one to detect the resonances down to zero transverse momentum.

The Muon Spectrometer is composed by three absorbers, a dipole magnet and two detector systems: the Muon Tracker (10 planes of Cathode Pad Chambers) and the Muon Trigger. The Muon Trigger works as an online trigger as an offline Muon Indentifier. Detailed descriptions of ALICE and its muon spectrometer are reported in [1] and [3].

## 2. The muon trigger system

The ALICE trigger system is composed of 4 planes (MT11, MT12, MT21, MT22) of Resistive Plate Chamber (RPC) detectors [4], arranged in two stations of 2 planes each. Each plane is in turn composed by 18 single gap RPCs. The two stations are placed at 16m and 17m from the interaction point, and arranged perpendicularly to the beam axis.

Each RPC is read out on both sides, allowing one to obtain two-dimensional position information. In this paper the vertical (horizontal) coordinate will be referred to as non-bending (bending), relatively to the dipole effect on charged particles. Each detection plane covers an area of about 5.5x6.5 $m^2$. The size of the RPC in the two stations is slightly different, to ensure equal angular acceptance. The strip size varies within each station to keep the occupancy of the detector constant. In the bending plane the pitches are about 1cm, 2cm and 4cm wide, while in the non-bending only 2cm and 4cm wide strips are installed.

The RPC detectors have low resistivity bakelite ($10^9 \div 10^{10}\Omega$m) electrodes [5] in order to enhance their rate capability. Both the cathodes and the gas gap are 2mm thick. The RPCs are operated with a gas mixture ($C_2H_2F_4$ 89.7%, i-$C_4H_{10}$ 10%, $SF_6$ 0.3%), optimized for a highly saturated avalanche operating mode [5].

Each plane is divided in 234 readout regions called Local Boards (LB). Since the last Long Shutdown two kinds of Front End Electronics (FEE) are in place:

- 71 chambers are still equipped with ADULT-FEE (working point ~10300 V) that allows for both avalanche and streamer working modes and provides no amplification [10].
- 1 chamber is equipped with a prototype of the FEERIC electronics (working point ~9700 V) which is being developed in view of RUNIII operations, as discussed elsewhere in these proceedings [6].

## 3. Running conditions in 2015

ALICE takes data in Pb–Pb collisions as well as in pp and p-Pb collisions, in order to provide a reference and a cold nuclear matter baseline. In Table 1 a summary of the different running conditions experienced so far is reported. In 2015 proton-proton data were taken at two different energies, plus a month of Pb–Pb collision data at 5.02 TeV.

| | |
|---|---|
| 2010 | Low luminosity pp and Pb–Pb collisions. Data taking and monitoring for the fine tuning of the system. |
| 2011 | • pp mainly @ $\sqrt{s} = 7\ TeV$ and some fills @ $\sqrt{s} = 2.76\ TeV$. $\mathcal{L}_{MAX} = 2\times10^{30} cm^{-2}s^{-1}$<br>• Pb–Pb @ $\sqrt{s_{NN}} = 2.76\ TeV$. $\mathcal{L}_{MAX} = 5\times10^{26} cm^{-2}s^{-1}$ |
| 2012 | • pp @ $\sqrt{s} = 8$ TeV. $\mathcal{L}_{MAX} = 7\times10^{30} cm^{-2}s^{-1}$ |
| 2013 | • p-Pb and Pb-p. $\mathcal{L}_{MAX} = 10^{29} cm^{-2}s^{-1}$<br>• pp @ $\sqrt{s_{NN}} = 2.76\ TeV$. $\mathcal{L}_{MAX} = 4\times10^{30} cm^{-2}s^{-1}$ |
| 2015 | • pp @ $\sqrt{s} = 13\ TeV$. $\mathcal{L}_{MAX} = 5\times10^{30} cm^{-2}s^{-1}$<br>• pp @ $\sqrt{s} = 5\ TeV$. $\mathcal{L}_{MAX} = 2\times10^{30} cm^{-2}s^{-1}$<br>• Pb–Pb @ $\sqrt{s_{NN}} = 5\ TeV$. $\mathcal{L}_{MAX} = 10^{27} cm^{-2}s^{-1}$ |

*Table 1: Summary of the ALICE running conditions in 2010-2015.*

## 4. Performance

### 4.1 Integrated charge 2010-2015

The integrated charge seen by each RPC so far has been determined using continuous measurements of the detector current, after subtraction of the dark current. As an example the results obtained for each RPC in detection plane MT21 are shown in Fig. 1. Such a plane contains the most exposed RPC of the system, which has an integrated charge of $\sim$9 mC/cm$^2$. The average integrated charge for all the 72 RPCs is $\sim 7\ \mathrm{mC/cm^2}$. Ageing tests have been performed up to $\sim 50\ \mathrm{mC/cm^2}$ with no worsening of the performances [5, 11].

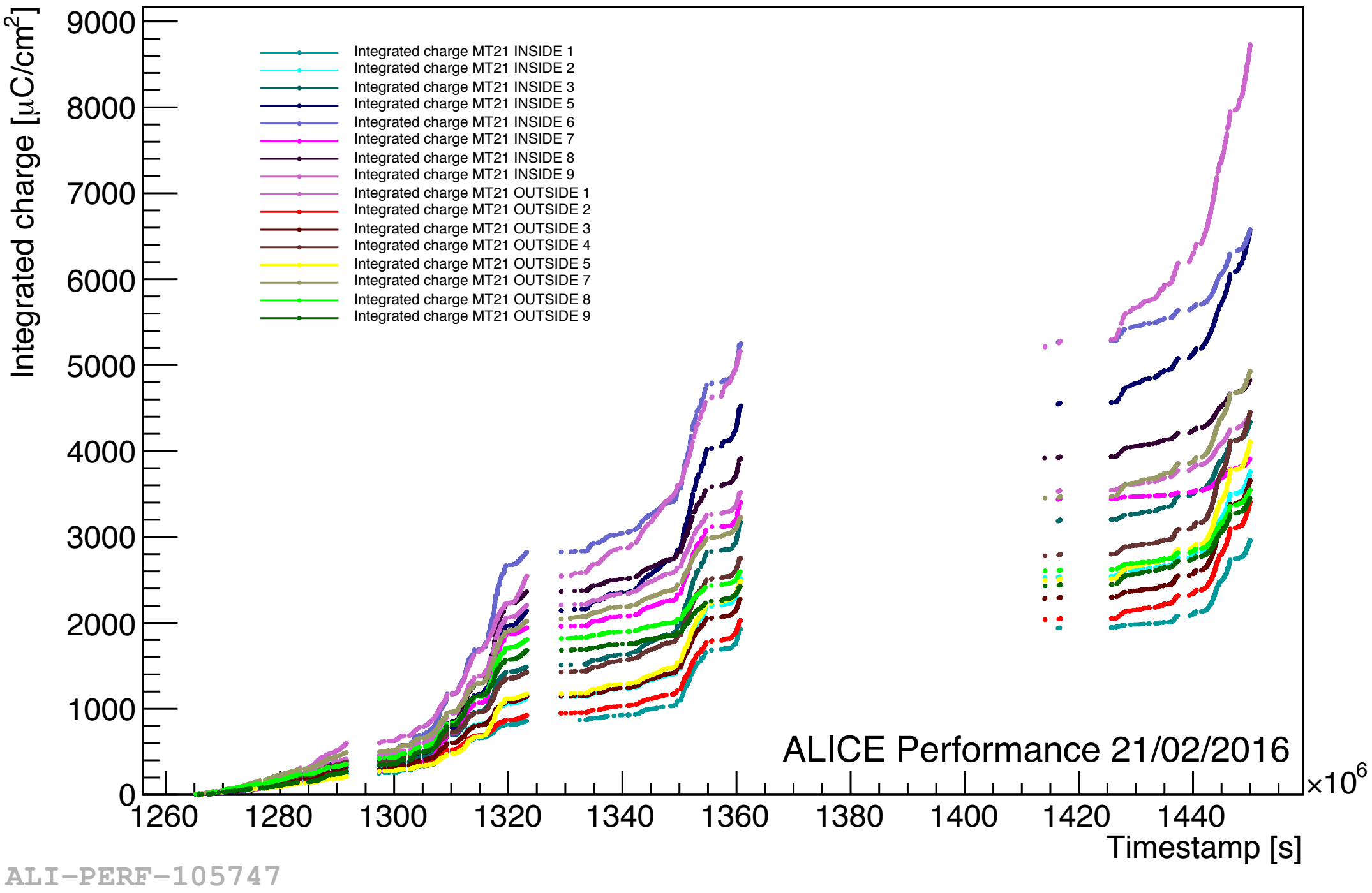

*Fig. 1 Integrated charge for the RPCs of MT21 in the period 2010-2015, as a function of time.*

### 4.2 Current distributions in different colliding systems

In order to compare the working conditions of RPCs for the two colliding systems of 2015, the current distributions in pp and Pb–Pb collisions are reported in Fig. 2. The results refer to the following luminosities ($\mathcal{L}$):

- pp collisions $\mathcal{L} = 5\times10^{30}\ \mathrm{cm^{-2}s^{-1}}$
- Pb–Pb collisions $\mathcal{L} = 10^{27}\ \mathrm{cm^{-2}s^{-1}}$

In spite of the very different luminosities very similar values are observed for the two systems. This is due to the much higher particle multiplicity of Pb–Pb with respect to pp collisions.

### 4.3 Charge per hit in different colliding systems

We now evaluate the average charge accumulated in the detectors per particle hit [12]. This value is the angular coefficient of the average current-average counting rate correlation, obtained by means of a linear fit. The computation has been performed for both pp and Pb–Pb collisions. The parameters of the two fits are observed to be compatible with each other and amount to $\sim 120\ \mathrm{pC/hit}$.

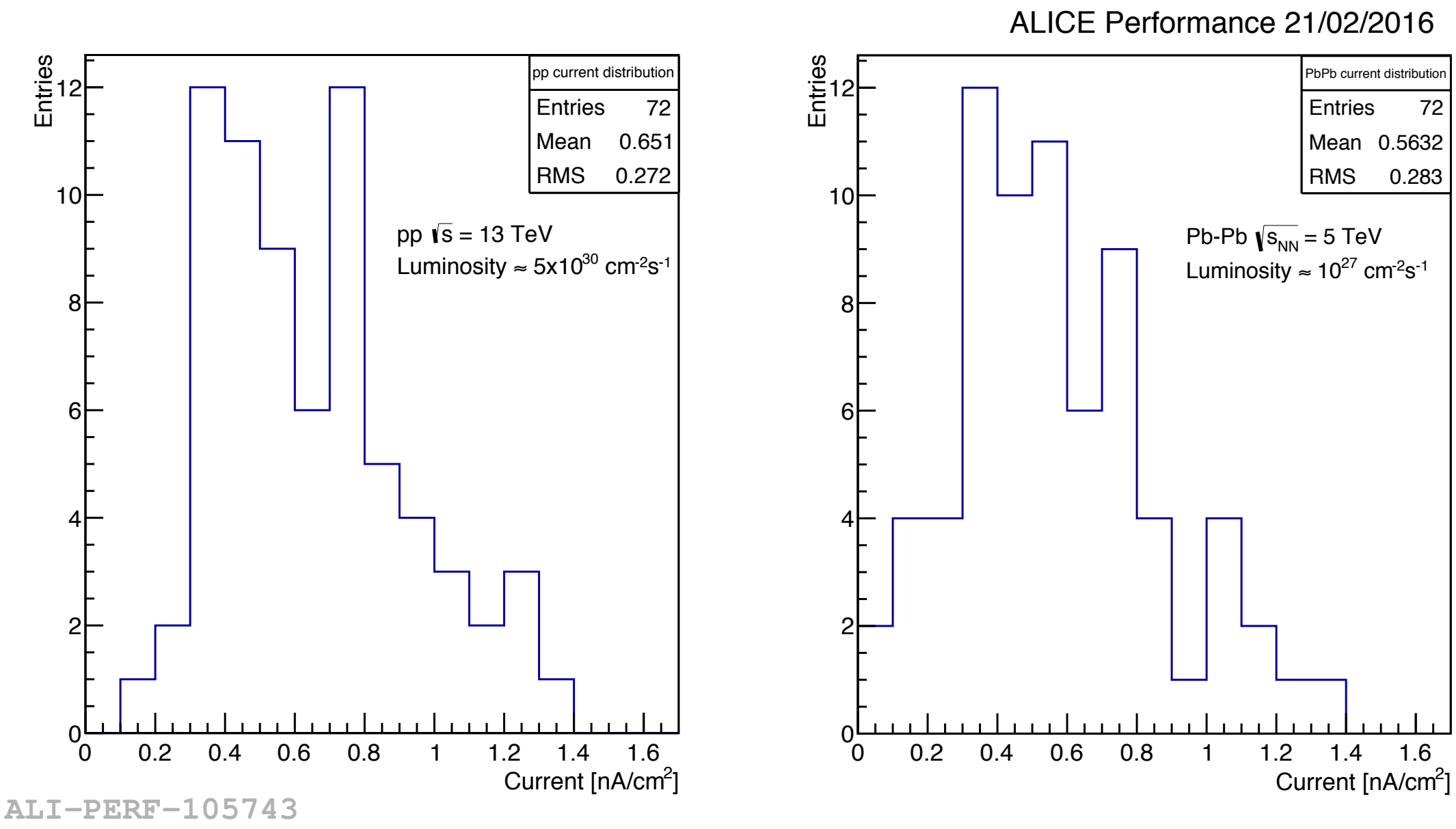

*Fig. 2 Total current distribution in pp (left) and Pb–Pb (right) collisions. One entry per RPC is reported (72 total per histogram).*

### 4.4 Dark rate and dark current

Fig. 3 shows the average dark rate and dark current of the muon trigger RPCs as a function of time. The dark rate is $< 0.1$ Hz/cm$^2$ and stable in time, although with some fluctuations. The dark current is $< 4.5$ µA and shows an increasing trend with time. The cause of this trend has been found to be related to few pathological RPCs whose behavior will have to be monitored during future data taking.

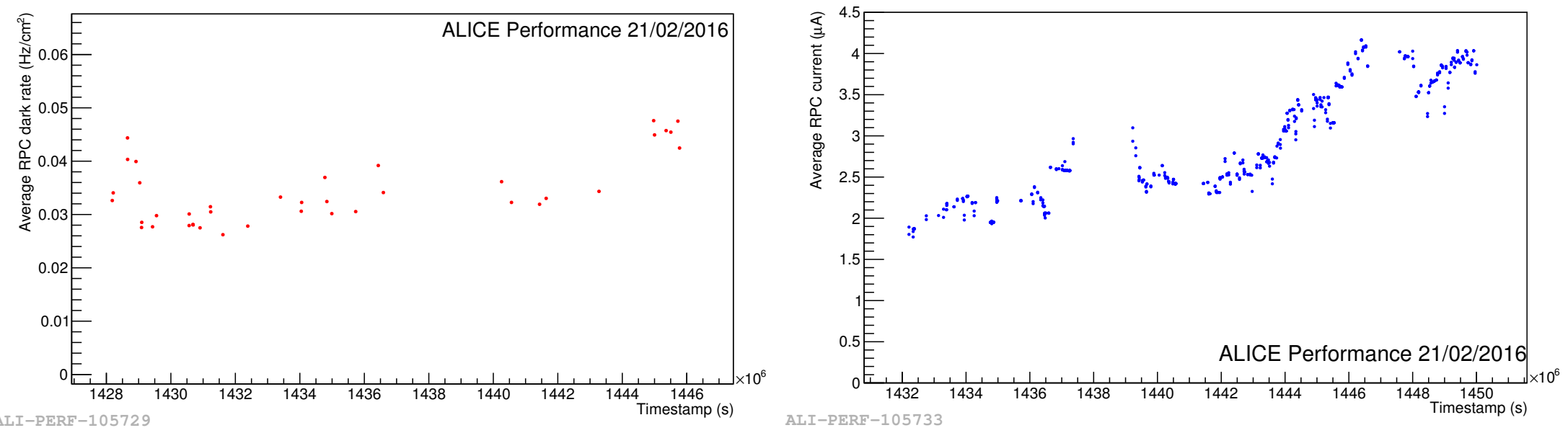

*Fig. 3 Average RPC dark rate (left) and average RPC dark current (right) as a function of time in 2015.*

### 4.5 Efficiency

The efficiency of the Muon Trigger RPCs has been computed by exploiting the redundancy of the trigger algorithm, which requires three out of four planes fired. The efficiency of a detection element in a given plane, can be determined by using the corresponding detection elements in the three other planes as an external trigger [13].

Fig. 4 (left) shows the efficiency of the 18 RPCs of MT21 in both the bending and non-bending planes. Fig. 4 (right) shows the average efficiency of MT21 as a function of time. Very similar results have been obtained for the other planes. The efficiency is typically >95%, uniform, and stable in time, independently of the running conditions.

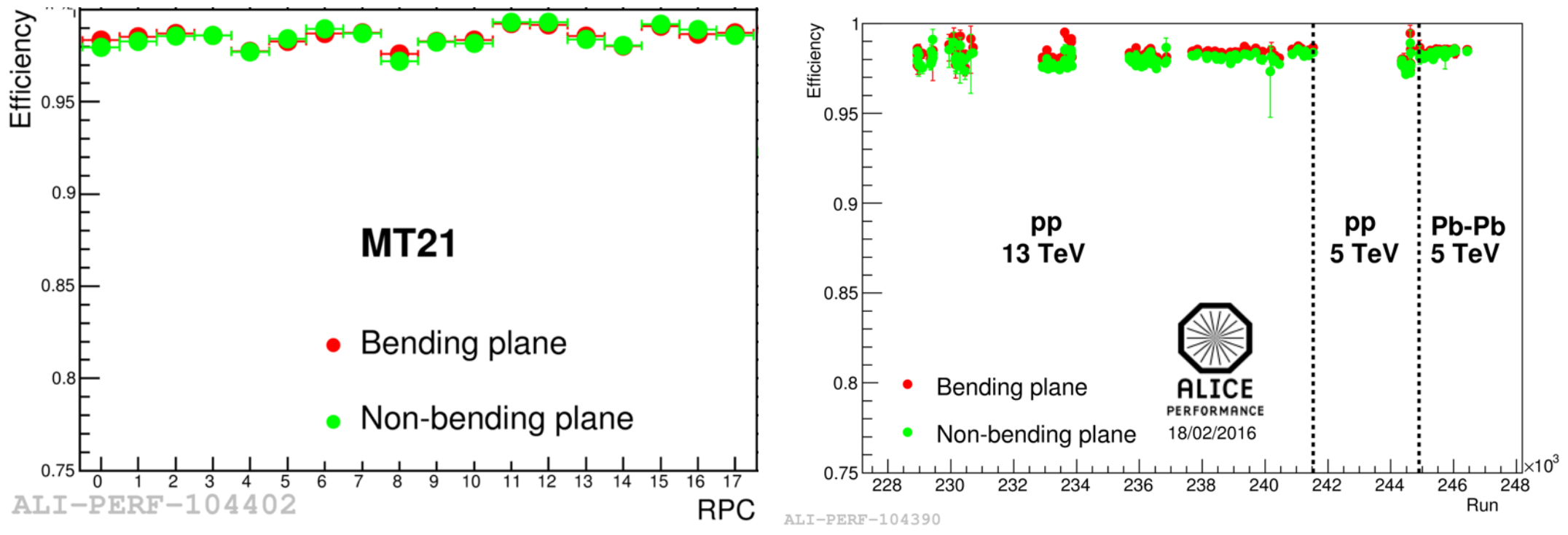

*Fig. 4 Left: Efficiency of the RPCs in MT21 as a function of the RPC ID. Right: Average efficiency of the RPCs of MT21 as a function of run number.*

### 4.6 Cluster size

In Fig. 5 the average cluster size for different strip pitches is shown for the different colliding systems explored since 2010. The trends appear to be stable over the whole period, except for a slight increase in 2015, that has to be further investigated.

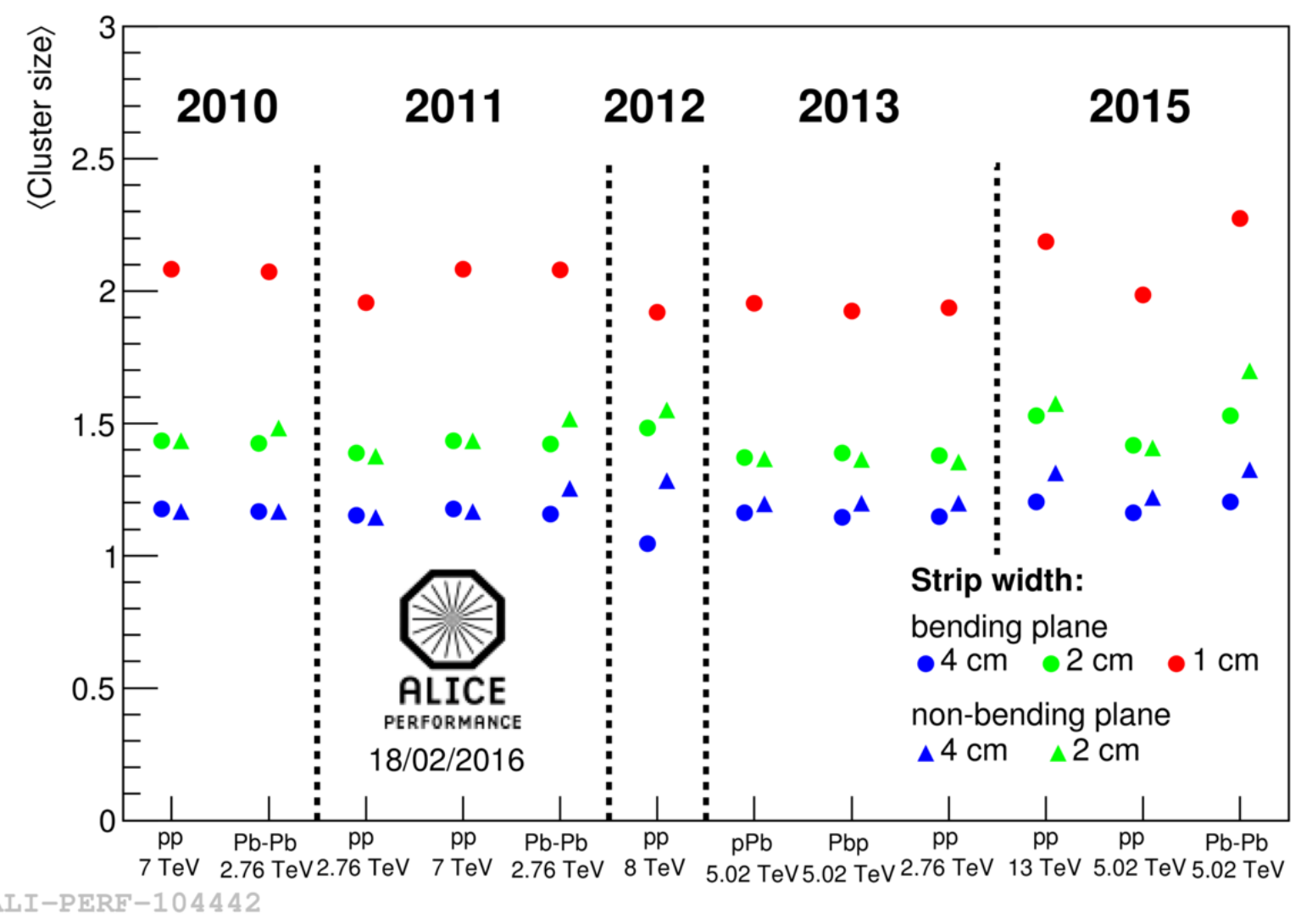

*Fig. 5 Average cluster size for different RPC strip pitches and colliding systems in 2010-2015.*

## 5. Conclusions

The RPCs of the Muon Spectrometer are working stably within specifications since 5 years, and 10 years after their installation in ALICE. The efficiency appears to be stable and >95% for all chambers. The measured cluster sizes are within specifications. A few RPCs show an increasing dark current, which has to be monitored.